\documentclass[11pt]{article}
\newcommand{\be}{\begin{eqnarray}}\newcommand{\beq}{\begin{equation}}
\newcommand{\ee}{\end{eqnarray}}\newcommand{\eeq}{\end{equation}}
\newcommand{\ep}{\epsilon}
\newcommand{\De}{\Delta}
\usepackage{graphicx} 
\title{
The effect of water-water hydrogen bonding on the hydrophobic hydration of macroscopic 
particles and its temperature dependence 
}
\author{Y. S. Djikaev\thanks{%Corresponding author. 
E-mail: idjikaev@buffalo.edu} 
\hspace{0.2cm}  \\ 
\\ Department of Chemical and Biological  Engineering, SUNY at Buffalo, \\ 
Buffalo, New York  14260 }
\date{ \hfill }
\renewcommand{\baselinestretch}{2}
\begin{document}
\renewcommand{\baselinestretch}{1}
\maketitle
\renewcommand{\baselinestretch}{1}
{\bf Abstract.} 
{\small 

A theoretical model for the effect of water hydrogen bonding on the  thermodynamics of
hydrophobic hydration is proposed as a combination of the classical density functional theory
with the recently developed  probabilistic approach to water hydrogen bonding in the vicinity
of a hydrophobic surface. The former allows one to determine  the distribution of water
molecules in the vicinity of a macroscopic hydrophobic particle and calculate the
thermodynamic quantities of hydrophobic hydration as well as their temperature dependence, 
whereas the latter allows one to implement the effect of the hydrogen bonding ability of 
water molecules on their interaction with the hydrophobic surface into the DFT formalism. This
effect arises because the number and energy of hydrogen bonds that a water  molecule forms 
near a hydrophobic surface differ from their bulk values. Such an alteration gives rise to a 
hydrogen bond contribution to the external potential field whereto a water molecule is
subjected in that vicinity. This contribution is shown to play  a dominant role in the
interaction of a water  molecule with the surface.  Our approach predicts that the free energy
of hydration of a planar hydrophobic surface in a model liquid water decreases with increasing
temperature in the range from 293 K to 333 K. This result is indirectly supported by  the
counter-intuitive experimental observation that under some conditions the hydration of a
molecular hydrophobe is entropically favorable as well as by the molecular dynamics
simulations predicting positive hydration entropy for  sufficiently large (nanoscale)
hydrophobes. 

} 

%{\em PACS}: 61.20.Gy; 61.20.Ne; 68.08.-p; 87.15.Fh \\
%{\em Keywords}: Hydrogen bonding; Hydrophobic surface; 
%Hydrophobic hydration; Free energy of hydrophobic hydration; Hydrophobic interactions
\renewcommand{\baselinestretch}{1} 
\newpage 
\section{Introduction} 
\renewcommand{\baselinestretch}{2}

Particles exhibiting resistance to being either  wetted by liquid water (case of
meso/macroscopic ones) or dissolved therein  (case of molecules) are generally referred to as
hydrophobic solutes.   The transfer of a hydrophobic solute into liquid  water is accompanied
by an increase in the free energy  of the system, which results from the structural
modifications of liquid water around the solute. This phenomenon is referred to as hydrophobic
hydration. When two solute particles are close enough to each other, the total volume of water
thus affected by both particles is smaller than when they are far apart. This gives rise to an
effective, solvent-mediated (often referred to as hydrophobic)  attraction between particles. 

It is believed that hydrogen bonding between neighboring water molecules$^{1-3}$ constitutes  a key
element of hydrophobic effects $^{4-7}$ (hydration and attraction)  which, in turn, play a crucial
role  in  many physical, chemical, and biological phenomena.$^{7-11}$ The development of predictive
models   capable of estimating the temperature and pressure dependence of hydrophobic phenomena  is
therefore quite important. 

In attempts to understand hydrophobic effects  at a fundamental level and to develop a generally
satisfactory theory of  hydrophobicity,$^{12,13}$  various mechanisms have been suggested$^{4-7,14-21}$
(mostly involving the  hydrogen bonding ability of water molecules). Despite many remaining
controversies, the dependence of hydrophobic phenomena  on the length scales of hydrophobic particles
involved appears to be out of contention.$^{22-25}$ 

Small hydrophobic molecules (whereof the linear size is comparable to that of a water molecule) can 
fit into the water hydrogen-bond network  without destroying any bonds.$^{15}$ One one hand, this 
results in a negligible enthalpy of  hydration; on the other hand, the presence of the solute is
thought to  constrain some degrees of freedom of the neighboring water molecules which should give 
rise to a negative entropy of hydration  proportional to the solute excluded volume. Consequently, the
hydration free energy is  positive and increases with temperature and solute excluded volume. 

Since the hydration of small hydrophobic molecules  is entropically ``driven", so is their
solvent-mediatated interaction.$^{12,13}$  At small enough separations, the two hydrophobic molecules 
affect fewer solvent molecules than when they are far apart. Therefore, bringing two  hydrophobic
molecules sufficiently close enough to each other should result  in a positive change of the entropy
and should thereby lower the free energy of the  solution (small enthalpy changes are neglected). 

Although attractive owing to its simplicity,  such a mechanism of small scale hydrophobicity appears to
be somewhat inaccurate.$^{12,13}$  Simulations$^{26,27}$ and theory$^{16}$ showed that two inert gas
molecules (such as argon)  would not be driven together to form a dimer; a solvent-separated pair would
be a more likely state than a contact pair. This leads to a surprising suggestion that the hydration of
small hydrophobic molecules is actually entropically favorable  (the entropy of the system increases)
which contradicts the conventional wisdom.  

The hydration of large hydrophobic particles is believed to occur via a different 
mechanism.$^{12,13,27-29}$  When inserted into liquid water,  such a particle breaks some hydrogen
bonds in its immediate vicinity.  This results in  a large positive  enthalpy of hydration and hence
in a free energy change proportional to the solute surface area (as opposed to being proportional to
the solute volume for small hydrophobes).

Thus, in contrast to entropically driven small-scale hydrophobicity,  the hydration of large
hydrophobic particles is expected to be enthalpically driven and  so is their hydrophobic interaction.
Fewer water hydrogen bonds have to be broken when two large hydrophobes are  ``in contact" than when
they are far from each other,  so there is a negative enthalpy change when  such particles aproach each
other from larger separations. The free energy change  (dominated by the enthalpy change) will be hence
negative and will constitute a thermodynamic driving force for their attraction. 

Among theoretical means for studying hydrophobic   phenomena (as well as many others including phase
transitions and phase equilibria), the methods of density functional theory$^{30,31}$ (DFT)  have
been particularly efficient. The DFT formalism  has been widely used for studying the density
profiles and thermodynamic properties of fluids  near rigid surfaces of various sizes, shapes, and
nature.$^{32,33}$ 

In DFT  the interaction of fluid
molecules with a foreign surface is usually  treated in the mean-field  approximation: every
fluid molecule is considered to be subjected to an external potential that arises due to its
pairwise interactions with the molecules of the impenetrable substrate.  This external
potential gives rise to a specific contribution to the free energy functional.  The
minimization of the free energy functional  with respect to the number density of fluid
molecules (as a function of  the spatial coordinate ${\bf r}$) provides their equilibrium
spatial distribution.  However, the effect of the impenetrable surface on the ability of
fluid (water) molecules to  form  hydrogen bonds near the surface had been ignored so far in
the conventional DFT formalism.

In the present work (that can be considered as a sequel of our recently published
paper$^{34}$), we attempt to fill in this gap and clarify some issues concerning the hydration
of large hydrophobic particles by combining  the density functional theory with the recently
developed probabilistic appraoch$^{35,36}$ to hydrogen  bonding between water molecules in the
vicinity of a foreign surface.$^{}$ This approach  provides an analytic expression for the
average number of hydrogen bonds  that a water molecule can form as a function of its distance
to the surface. Knowing this expression, one can implement the effect of the hydrogen bonding
of water molecules on their interaction with the hydrophobic surface into DFT, which is then
employed to determine  the distribution of water  molecules near a macroscopic hydrophobic
particle and to calculate the thermodynamic quantities of hydrophobic hydration and their
temperature dependence. 

\section{The outline of a probabilistic approach to water--water 
hydrogen bonding near a hydrophobic surface} 

Let us first briefly describe the probabilistic hydrogen bond (PHB) model$^{35,36}$ for the
hydrogen bonding ability of water molecules. It considers  a water molecule, whereof the
location is determined by its center,   to have four hydrogen-bonding (hb) arms (each capable
of forming a single hydrogen bond) of  rigid and symmetric (tetrahedral) configuration with
the inter-arm angles  $\alpha=109.47^\circ$.  Each hb-arm can  adopt a continuum of
orientations.  For a water molecule to form a hydrogen bond with  another molecule, it is
necessary  that  the tip of any of its hb-arms coincide with  the second molecule. The length
of an hb-arm thus equals the  length $\eta$ of a hydrogen bond. 
%(The sufficient condition for a hydrogen bond between two
%water  molecules to form would  require that any hb-arm of one molecule roughly coincide with
%any hb-arm of another.) 
%Our model does not recur to this requirement).

The hydrogen bond length $\eta$ is assumed to be independent of whether the molecules  are in
the bulk or near a hydrophobic surface.  The characteristic  length $\widetilde{\eta}$  of
pairwise interactions between water and molecules  constituting the
substrate (flat and  large enough to neglect edge effects, with its location 
determined by the loci of the  centers of its outermost, surface
molecules)  plays a simple role in the 
hydrogen bond contribution to hydration or  hydrophobic interaction.$^{35,36}$ 
It  ($\widetilde{\eta}$) only determines 
the  reference point for measuring the distance between water molecule and substrate, so it
will be set equal to $\eta$.  
%Namely, if
%the water and substrate  molecules were treated as hard  spheres, then  this length would
%just determine the minimal distance between water  molecule and substrate such that a
%planar layer of thickness $\widetilde{\eta}$ adjacent to the plate would be excluded to water
%molecules (In reality, water molecules are not  hard spheres and they do approach the plate
%to distances somewhat closer than $\widetilde{\eta}$).  
%The actual value of 
%$\widetilde{\eta}$ plays a simple role in the model. 
%It ($\widetilde{\eta}$) only determines
%the  reference point for measuring the distance between water molecule and substrate
%One can thus assume $\widetilde{\eta}$ to be roughly equal to $\eta$. 

Consider a ``boundary" water molecule (BWM) in  the vicinity of a hydrophobic substrate S
(immersed in liquid water) at a distance $x$ therefrom. Such a molecule  forms a smaller
number  of hydrogen bonds (hereafter referred to as ``boundary hydrogen bonds")  than in bulk
water  because the hydrophobic surface restricts  the configurational space available to
other water molecules that are  necessary for a BWM to form hydrogen bonds. 
The actual number of hydrogen bonds, that a particular BWM can form,  depends on both its
location and its orientation.  A probabilistic hydrogen bond model allows one to obtain an
analytic expression for the average number of bonds that a BWM  can form as a function of its
distance to the surface  (``average" with respect to all possible orientations of the water
molecule).$^{35,36}$  

Note that a 
boundary hydrogen bond (BHB), involving at least one boundary water molecule, 
may be slightly altered energetically compared to the bulk one. 
Such alteration is still a subject of contention$^{37}$ as different authors suggest 
opposite effects,$^{}$ i.e., both enhancement 
$^{18,38}$ and weakening$^{3a}$ of the boundary hydrogen bonds.
In the PHB approach, there is no restriction on the energy of a bulk
(water-water) hydrogen bond, $\ep_b<0$, 
so that the approach is valid independent of whether
$\ep_b<\ep_s$ or $\ep_b=\ep_s$, or $\ep_b>\ep_s$, where $\ep_s<0$ is the energy of a BHB. 

\subsection{The average number of BHB's per water molecule} 

Let us choose a Cartesian coordinate system so that its $x$-axis is normal to the plate $P$
located at $x=0$.  Denote the number of hydrogen bonds per bulk water molecule by $n_b$ and
the {\em average} number of hydrogen bonds per BWM by $n_s$. The latter is a  function of
distance $x$  between the water molecule  and the hydrophobic surface (assumed to be smooth on
a molecular scale): $n_s\equiv n_s(x)$.  If $x$ is larger than $2\eta$, the number of hydrogen
bonds that the molecule can form  is not affected by the presence of the surface. Therefore,
$n_s(x)=n_b$ for $x\ge 2\eta$.  On the other hand, the function $n_s(x)$ attains its minimum
at the minimal distance between the water molecule and the plate, i.e., 
%$n_s(x)=n_s^{min}$ 
at $x=\eta$, 
because at this distance the configurational space available for 
the neighboring water molecules (to form a bond with the selected one) is restricted 
(compared to the bulk water) by the plate the most. 
The layer of thickness $\eta$ from $x=\eta$ to 
$x=2\eta$ is referred to as the surface hydration layer (SHL). 

The function $n_s=n_s(x)$ can be shown$^{35,36}$ to have the form  
\beq n_s=k_1b_1+k_2b_1^2+k_3b_1^3+k_4b_1^4.\eeq 
where $b_1$ is the probability that one of the hb-arms (of a bulk water molecule) can form a 
hydrogen bond and $k_1\equiv k_1(x),\;k_2\equiv k_2(x),\;
k_3\equiv k_3(x)$, and 
$k_4\equiv k_4(x)$ are coefficient-functions that can be 
evaluated by using geometric considerations (with their dependence on the BWM orientations being
averaged). The functions $k_1(x),k_2(x),k_3(x)$, and $k_4(x)$ are presented in Figure 2a of ref.36. 
They all become equal to $1$ at $x\ge 1$ where eq.(1) reduces to its bulk analog, 
$n_b=b_1+b_1^2+b_1^3+ b_1^4$ (see ref.39). Since 
experimental data on $n_b$ (and even its temperature dependence) are
readily available, the latter equation allows one to determine 
the probability $b_1$ as its positive solution satisfying the condition $0<b_1<1$. 
Thus, equation (1) provides an efficient pathway to $n_s$ as a function of $x$. 
Figure 2b in ref.36 presents this function for a 
hydrophobic flat surface immersed in water at temperature $T=293.15$ K, which corresponds
to  $n_b=3.65$ hence $b_1=0.963707$.  
%As expected, the average
%number of hydrogen bonds per water molecule monotonically increases from its minimal value at
%$x=\eta$  (the closest possible distance to the plate) to its maximum bulk value $n_b$ at
%$x=2\eta$. 

The above expression for $n_s$ takes into account  the  constraint that some  orientations
of  the  hb-arms of a BWM cannot lead to the formation of hydrogen bonds  because of the
proximity to the  hydrophobic  particle. The severity of this constraint depends on the
distance of the BWM to the surface, hence the $x$-dependence of $k_1, k_{2}, k_{3},$ and
$k_{4}$.  It assumes that  the intrinsic hydrogen-bonding ability of a BWM (i.e., the
tetrahedral configuration of its hb-arms and their lengths and energies) are unaffected by 
its proximity to the hydrophobic surface so that the  latter only restricts the
configurational space available to  other water molecules  necessary for this BWM to form
hydrogen bonds.  

\subsection{Hydrogen bond contribution to the interaction of a water molecule with a 
hydrophobic plate}

Knowing the function $n_s(x)$, one can examine the effect of water hydrogen bonding  on the
hydration  of hydrophobic (and even composite) particles as well as on their solvent-mediated
interaction.  For example, let us derive an expression for  $U_{\mbox{\tiny
ext}}^{\mbox{\tiny hb}}\equiv U_{\mbox{\tiny ext}}^{\mbox{\tiny hb}}(x)$,  water-water
hydrogen bond contribution to the total  external potential field  $U_{\mbox{\tiny
ext}}^{\mbox{\tiny }}\equiv U_{\mbox{\tiny ext}}^{\mbox{\tiny }}(x)$  whereto a water
molecule is subjected in the vicinity of a hydrophobic surface. The latter is needed for the
application of DFT methods  to the thermodynamics of hydrophobic phenomena. 

The hydrogen bond contribution 
$U_{\mbox{\mbox{\tiny ext}}}^{\mbox{\mbox{\tiny hb}}}$ is due to 
the deviation of $n_s$ from $n_b$ (see refs.35 and 36) and 
the (possible) deviation of $\ep_s$ from $\ep_b$. 
It can be determined as 
\beq U_{\mbox{\tiny ext}}^{\mbox{\tiny hb}}\equiv U_{\mbox{\tiny ext}}^{\mbox{\tiny hb}}(x)=
\ep_s(x)n_s(x)-\ep_bn_b \;\;\;\;\;\;\;\;\;(\eta\le x<\infty).\eeq
The first term on the RHS of this equation represents the total energy of hydrogen bonds 
of a water molecule at a distance $x$ from 
the surface, whereas the second term is the energy of its hydrogen bonds in bulk (i.e., at
$x\rightarrow \infty$. 
Note that the dependence 
of $U_{\mbox{\tiny ext}}^{\mbox{\tiny hb}}$ on $x$ may be due not only to the function
$n_s(x)$, but also to the $x$-dependence of the hydrogen bond energy in the vicinity of the 
hydrophobic surface, $\ep_s\equiv \ep_s(x)$. In the PHB model 
$n_s(x)=n_b$ for $x\ge 2\eta$  hence it is reasonable to assume
that $\ep_s(x)=\ep_b$ for $x\ge 2\eta$ as well. Thus, 
$U_{\mbox{\tiny ext}}^{\mbox{\tiny hb}}(x)$ is a very
short-ranged function of $x$, such that  $U_{\mbox{\tiny ext}}^{\mbox{\tiny hb}}(x)=0$ for 
$x\ge 2\eta$.

\section{The outline of the methods of density functional theory}

The effect of water-water hydrogen bonding on the density profile of (liquid) 
water molecules in the vicinity of a hydrophobic surface can be now examined by using
DFT.$^{30-33}$ 
In this formalism, the grand thermodynamic potential $\Omega$ of a
nonuniform single component fluid, subjected to an external potential $U_{\mbox{\tiny ext}}$, 
can be represented as a functional of the number density $\rho(\bf{r})$ of fluid molecules 
%\be \Omega[\rho(\bf{r})]&=&\int_{V}d{\bf r}\,f_h(\rho({\bf r}))\nonumber\\ 
%&+& \frac1{2}\int\int d{\bf r} d{\bf r'}\,\rho({\bf r})\rho({\bf r'}) 
%\phi_{\mbox{\tiny at}}(|\bf{r}-\bf{r'}|)
%+\int d {\bf r}\, U_{\mbox{\tiny ext}}({\bf r})\rho({\bf r})-
%\mu\int d{\bf r}\, \rho({\bf r}),\ee
\beq \Omega[\rho({\bf r})]=\int_{V}d{\bf r}\,f_h(\rho({\bf r}))
+ \frac1{2}\int\int d{\bf r} d{\bf r'}\,\rho({\bf r})\rho({\bf r'}) 
\phi_{\mbox{\tiny at}}(|{\bf r}-{\bf r'}|)
+\int d {\bf r}\, U_{\mbox{\tiny ext}}({\bf r})\rho({\bf r})-
\mu\int d{\bf r}\, \rho({\bf r}),\eeq
where $V$ is the volume of the system, $\mu$ is the equilibrium chemical potential, and 
$\phi_{\mbox{\tiny at}}(|\bf{r}-\bf{r'}|)$ is the attractive part of the interaction
potential between two fluid molecules located at ${\bf r}$ and ${\bf r'}$. In this
expression, the contribution to the free energy due to the short range repulsive interactions
(the first term on the RHS of the equation) is modeled by the hard sphere free energy in a
local density approxiamtion (LDA), with $f_h$ at ${\bf r}$ being the Helmholtz free energy
density of a hard sphere fluid of uniform density equal to $\rho({\bf r})$.   
The longer ranged attractive interactions are treated in a mean-field (van der Waals)
approximation and represented by the second term on the RHS of eq.(3). (Note that the  LDA
neglects short-ranged correlations which leads to the absence of oscillations in the density
profile of a fluid near a hard wall. Although more accurate, nonlocal  approximations 
are also available,$^{31,40,41}$  we preferred the LDA to ensure the transparency of presentation and
to put the emphasis on the idea of combining the DFT methods  with the PHB model).

In an open thermodynamic system   under conditions of constant chemical potential $\mu$,
volume $V$,  and temperature $T$ (grand canonical ensemble), the  equilibrium density profile
is obtained by minimizing the functional  $\Omega[\rho(\bf{r})]$ with respect to $\rho({\bf
r})$, i.e., by solving the Euler-Lagrange equation $\delta \Omega/\delta\rho=0$, which takes
the form 
\beq \mu=\mu_h(\rho({\bf r}))
+\int_V d{\bf r'}\,\rho({\bf r'}) \phi_{\mbox{\tiny at}}(|\bf{r}-\bf{r'}|)
+U_{\mbox{\tiny ext}}({\bf r}),\eeq
where $\mu_h(\rho)\equiv d f_h(\rho)/d\rho$ is the chemical potential of the uniform reference 
(hard sphere) fluid of density $\rho$. The substitution of the equilibrium density profile 
$\rho({\bf r})$ into eq.(3) provides the grand thermodynamic potential $\Omega$ of the fluid. 

As already mentioned, the term $U_{\mbox{\tiny ext}}$ on the RHS of eq.(4) (and the
corresponding term on the RHS of eq.(3))  had been conventionally meant to represent the
external potential exerted by all the molecules constituting the hydrophobic substrate on a
fluid molecule. Various models for the external potential were designed to take into account
pairwise interactions of a fluid  molecule with the molecules of the substrate$^{32,33}$ as
well as the effect of the latter on the pairwise  interactions between fluid molecules
themselves.$^{42}$ The contribution of these (pairwise) effects into
$U_{\mbox{\tiny ext}}$ will be denoted by  $U_{\mbox{\tiny ext}}^{\mbox{\tiny pw}}$ to
distinguish it from the hydrogen bond contribution, $U_{\mbox{\tiny
ext}}^{\mbox{\tiny hb}}$. Thus, the overall external  potential whereto a water molecule is
subjected near a hydrophobic surface can be represented as 
\beq U_{\mbox{\tiny ext}}({\bf r})=U_{\mbox{\tiny ext}}^{\mbox{\tiny pw}}({\bf r})+
U_{\mbox{\tiny ext}}^{\mbox{\tiny hb}}({\bf r}).\eeq

In a closed thermodynamic system with constant number of molecules $N$, volume $V$,  and
temperature $T$ (canonical ensemble), the chemical potential $\mu$, appearing in eqs.(3) and
(4), is not known in advance. Instead, it plays the role of a Lagrange multiplier
corresponding to the constraint of fixed number of molecules in the system:
$$ N=\int_{V}d{\bf r}\,\rho(r). $$
This equation can be used to determine the Lagrange multiplier $\mu$, i.e., chemical potential
in the system, as follows (see, e.g., ref.43). Introducing the ``configurational" part of the
hard sphere  chemical potential as $\widetilde{\mu_h}(\rho({\bf r}))\equiv \mu_h(\rho({\bf
r})) -k_BT\ln\rho({\bf r})$, one can rewrite equation (3) in the form
$$ \rho({\bf r})=\exp\Big\{\frac1{k_BT}\Big[\mu-U_{ \mbox{\tiny ext} }({\bf r})-\widetilde{\mu}(\rho({\bf r'}))-\int d{\bf r'}\rho({\bf
r'})\phi_{\mbox{\tiny at}}(|{\bf r}-{\bf r'}|)\Big]\Big\}. $$
Integrating this equation over the volume of the system and using the constraint on $N$, one
obtains 
$$ \mu=k_BT\ln N -k_BT\ln\left( \int d{\bf r}\exp
\Big\{\frac1{k_BT}\Big[-U_{ \mbox{\tiny ext} }({\bf r})-\widetilde{\mu}(\rho({\bf r'}))-\int d{\bf r'}\rho({\bf
r'})\phi_{\mbox{\tiny at}}(|{\bf r}-{\bf r'}|)\Big]\Big\}\right),$$ 
whereof the substitution into eq.(3) yields the Euler-Lagrange
equation for the density profile in the canonical ensemble: 
\be 
\mu_h(\rho({\bf r}))&=&k_BT\ln N -k_BT\ln\left( \int d{\bf r}\exp
\Big\{\frac1{k_BT}\Big[-U_{ \mbox{\tiny ext} }({\bf r})-\widetilde{\mu}(\rho({\bf r'}))-\int d{\bf r'}\rho({\bf
r'})\phi_{\mbox{\tiny at}}(|{\bf r}-{\bf r'}|)\Big]\Big\}\right)\nonumber \\
&-& U_{ \mbox{\tiny ext} }({\bf r}) -
\int d{\bf r'}\rho({\bf r'})\phi_{\mbox{\tiny at}}(|{\bf r}-{\bf r'}|),
\ee
The Helholtz free energy $F$ of the canonical ensemble can then be 
obtained by substituting the
solution of eq.(6) into the corresponding functional of $\rho({\bf r})$:
\beq F[\rho({\bf r})]=\int_{V}d{\bf r}\,f_h(\rho({\bf r}))
+ \frac1{2}\int\int d{\bf r} d{\bf r'}\,\rho({\bf r})\rho({\bf r'}) 
\phi_{\mbox{\tiny at}}(|{\bf r}-{\bf r'}|)
+\int d {\bf r}\, U_{\mbox{\tiny ext}}({\bf r})\rho({\bf r}),\eeq

In the particular case of (fluid) water near a flat hydrophobic surface, one can use the
planar symmetry of the system and choose the Cartesian coordinates so that the surface is
located in the $y-z$ plane at $x=0$ with the molecules of the fluid occupying the
``half-space" $x>0$. The  eqilibrium density profile obtained from eqs.(4) or (6) is then a
function of a single  variable $x$, i.e., $\rho({\bf r})=\rho(x)$. 

\section{Free energy of hydration and its temperature dependence}

In a canonical ensemble, 
the free energy $\De F_{\mbox{\tiny }}$ 
of hydration of a hydrophobic particle can be determined as the difference 
\beq \De F_{\mbox{\tiny }}=F-F_0,\eeq 
where $F$ and $F_0$ are the Helmholtz free energies of the system (liquid water) with and without 
a hydrophobic particle therein, respectively.  
Likewise, the free energy of hydrophobic hydration in a grand canonical ensemble can be
determined as 
\beq \De \Omega_{\mbox{\tiny }}=\Omega-\Omega_0,\eeq 
where $\Omega$ and $\Omega_0$ are the values of the grand thermodynamic potential 
of the system (liquid water) with and without a hydrophobic particle therein, respectively.

Knowing the free energy of hydrophobic hydration, one can find $\Phi_{\mbox{\tiny S}}$ and  
$\Phi_{\mbox{\tiny E}}$, the entropic and energetic contributions to $\De F$, as  
\beq \Phi_{\mbox{\tiny S}}\equiv-T\De S_{\mbox{\tiny }}=T(\partial \De F_{\mbox{\tiny }}/\partial T)_{\mbox{\tiny
V,N}},\;\;\;\;\; \Phi_{\mbox{\tiny E}}\equiv \De E_{\mbox{\tiny }}=(\partial (\De F_{\mbox{\tiny }}/T)/\partial
(1/T))_{\mbox{\tiny V,N}},
\eeq  
respectively, such that  $\De F_{\mbox{\tiny }}=\Phi_{\mbox{\tiny E}}+\Phi_{\mbox{\tiny S}}$ 
(in eq.(10) the subscripts of the partial derivatives
indicate the thermodynamic variables held constant upon taking the derivatives). 
Clearly, for the decomposition of $\De F_{\mbox{\tiny }}$ into energetic and entropic
components   it is necessary to know its temperature dependence. (Note that a similar
decomposition can be carried out for $\De \Omega_{\mbox{\tiny }}$.)

In the combined PHB/DFT-based model presented above, the temperature dependence of 
$\De F_{\mbox{\tiny }}$ contains a contribution from the temperature dependence of  
$U_{\mbox{\tiny ext}}^{\mbox{\tiny hb}}$, 
hydrogen bond contribution to the overall external field exerted by the hydrophobic surface on
water molecules in its vicinity. As clear from eq.(2), 
the dependence of $U_{\mbox{\tiny ext}}^{\mbox{\tiny hb}}$ on $T$ is 
due to the temperature dependence of four quantities: $n_s,\,n_b,\,
\ep_s$, and $\ep_b$. The functions  $\ep_b\equiv \ep_b(T)$ and
$n_b\equiv n_b(T)$ are either readily available or can be constructed on the basis of available
data.$^{}$ On the other hand, $n_s$ is unambiguously related to $n_b$, hence its  dependence on $T$
can be considered to be known as well. Finally, it is reasonable to assume that,  whether in the
bulk or in the surface hydration layer, the energy of a hydrogen bond depends on temperature in
such a way that the ratio $\ep_s(T)/\ep_b(T)$ is independent of $T$. One can thus consider 
$ U_{\mbox{\tiny ext}}^{\mbox{\tiny hb}}$ 
to be a known function of not only $x$ but also $T$, $U_{\mbox{\tiny ext}}^{\mbox{\tiny
hb}}=U_{\mbox{\tiny ext}}^{\mbox{\tiny hb}}(x,T)$. This allows one to numerically determine the
temperature dependence of $\De F_{\mbox{\tiny }}$ and to subsequently use interpolation procedure to
find an accurate analytical fit thereof which then can be used in eq.(10).

\section{Numerical evaluations}

In order to illustrate the 
above model with numerical calculations, we have considered the hydration of a flat,
macroscopic hydrophobic surface in a model liquid water. 
The pairwise interactions between two water molecules 1 and 2 were 
modeled with the Lennard-Jones (LJ) potential, 
$$\phi_{\mbox{\tiny ww}}=4\ep_{\mbox{\tiny ww}}
\Big[\Big(\frac{d}{r_{12}}\Big)^{12}-\Big(\frac{d}{r_{12}}\Big)^{6}\Big],$$
where $r_{12}$ is the distance between molecules 1 and 2,  $\ep_{\mbox{\tiny ww}}$ is the
energy parameter and $d$ is  the diameter of a model molecule. The parameter $\ep_{\mbox{\tiny
ww}}$ was adjusted to be $3.79\times 10^{-14}$, which differs  from its values used in the
computer  simulations (Monte Carlo or molecular dynamics) of various water models,  in which
$\ep_{\mbox{\tiny ww}}$ ranges$^{3b}$ from  $5.31\times 10^{-15}$ erg (ST2 model) to
$1.47\times 10^{-14}$ erg (SWM4-NDPmodel) to  $2.54\times 10^{-13}$ erg (SSD model). 
Such a modification was needed to ensure that  the phase diagram of model water more or less
resembles that of real water. In  the DFT formalism this can be achieved only by adjusting the
single intermolecular potential  describing water-water interactions ,  whereas in computer
simulations water-water  interactions are usually described by the combination of LJ and
electrostatic potentials;  hence the difference in the energy parameters of the respective LJ
potentials. 
The parameter $d$ of the LJ potential also has different values in different 
water models$^{3b}$ in the range from $3.02$ {\AA} (SSD model) to 
$3.18$ {\AA} (SWM4-NDP model). On the other hand, the length $\eta$ of the hydrogen bond (i.e., the
distance between the oxygen atoms of two hydrogen-bonded water molecules) is reported$^{3b}$ 
to be about $2.98$ {\AA}. Since $d$'s (of various water models) and $\eta$ are so close to 
each other, we assumed for our model $d\simeq \eta$.
 
To find the equilibrium density profile of (model) water molecules in the vicinity of the hydrophobic
surface, it is necessary to solve eq.(4) using, e.g., an iterative procedure.$^{33}$ 
Namely, the density profile $\rho_{i}(x)$at the $i^{\mbox{\tiny th}}$ 
iteration is found from the previous one, $\rho_{i-1}(x)$ via 
\beq \mu_h(\rho_{i}(x))=\mu
-\int_V d{\bf r'}\,\rho_{i-1}(x) \phi_{\mbox{\tiny at}}(|{\bf r}-{\bf r'}|)
-U_{\mbox{\tiny ext}}(x). \eeq 
A similar iterative procedure can be used to solve equation (6), 
\be 
\mu_h(\rho_{i}(x))&=&k_BT\ln N -k_BT\ln\left( \int d{\bf r}\exp
\Big\{\Big[-U_{ \mbox{\tiny ext} }(x)-\widetilde{\mu}(\rho_{i-1}(x'))-
\int d{\bf r'}\rho_{i-1}(x')\phi_{\mbox{\tiny at}}(|{\bf r}-{\bf r'}|)\Big]\Big\}\right)\nonumber \\
&-& U_{ \mbox{\tiny ext} }(x) -
\int d{\bf r'}\rho_{i-1}(x')\phi_{\mbox{\tiny at}}(|\bf{r}-\bf{r'}|).
\ee

For the chemical potential $\mu_h(\rho)\equiv\mu_h(\rho,T)$ 
of a hard sphere fluid we have adopted the well-known
Carnaham-Starling approximation$^{32,33,44}$ 
$$\mu_h(\rho,T)=k_BT\Big(\ln(\Lambda^3\rho)+\xi\frac{8-9\xi+3\xi^2}{(1-\xi)^3}\Big) $$ 
where $\xi=(\pi d^3/6)\rho$ and 
$\Lambda=(h^2/2\pi m k_BT)^{1/2}$ is the thermal de Broglie wavelength 
of a model molecule of mass $m$ 
(with $h$ and $k_B$ being Planck's and 
Boltzmann's constants, respectively). 
Since $\mu_h$ is a single-valued (monotonically increasing) 
function of $\rho$, one can extract $\rho_i(x)$ from the LHS of eq.(11) or (12) and continue iterations. 
%Note that the pairwise interactions of water molecules were modeled by using the Lennard-Jones 
%potential, 
The attractive part $\phi_{\mbox{\tiny at}}$ of the pairwise water-water interactions was modeled
by using a well-known perturbation scheme:$^{45}$
$$ \phi_{\mbox{\tiny at}}(r_{12})=\left\{\begin{array}{ll} 
-\ep_{\mbox{\tiny ww}} & (r_{12}<2^{1/6}\eta),\\
\phi_{\mbox{\tiny ww}}, & 
(r_{12}>2^{1/6}\eta). \end{array} \right.  
$$ 
 
%potential $\phi_{\mbox{\tiny ww}}$ was 
%adopted 
%according to the perturbation scheme 
%as suggested in ref.28.
%The attractive part of the
%water-water interactions was taken to be
%\beq U_{\mbox{\tiny ext}}^{\mbox{\tiny pw}}(x)=\left\{\begin{array}{ll} 
%\infty & (x<\eta),\\
%-\ep_{\mbox{\tiny sw}}\exp[-\lambda_{\mbox{\tiny sw}}\,(x-\eta)] & (x>\eta),
%\end{array} \right.\eeq 

Figure 1 presents the typical behaviour of $U_{\mbox{\tiny ext}}$ and its components. 
The thick dashed curve shows $U_{\mbox{\tiny ext}}$, while the lower and upper 
thin solid curves are for $U_{\mbox{\tiny ext}}^{\mbox{\tiny pw}}$ and 
$U_{\mbox{\tiny ext}}^{\mbox{\tiny hb}}$,
respectively. The function $U_{\mbox{\tiny ext}}^{\mbox{\tiny pw}}(x)$ 
was modeled as suggested in refs.32 and 33, 
\beq U_{\mbox{\tiny ext}}^{\mbox{\tiny pw}}(x)=\left\{\begin{array}{ll} 
\infty & (x<\eta),\\
-\ep_{\mbox{\tiny sw}}\exp[-\lambda_{\mbox{\tiny sw}}\,(x-\eta)] & (x>\eta),
\end{array} \right.\eeq
where the energy parameter $\ep_{\mbox{\tiny sw}}$ was taken to be equal to 
$k^{\mbox{\tiny }}_{\mbox{\tiny p}}\ep_{\mbox{\tiny ww}}$ with $k_{\mbox{\tiny p}}=2.1$
and the inverse length parameter $\lambda_{\mbox{\tiny sw}}$ was set equal to $1/\eta$ 
(the ratio 
$k^{\mbox{\tiny }}_{\mbox{\tiny p}}\equiv \ep_{\mbox{\tiny sw}}/\ep_{\mbox{\tiny ww}}$  
characterizes the degree of hydrophobicity of the surface as  
the energy of a water molecule  attraction to the  surface at the distance $\eta$ between 
them relative
to  $-\ep_{\mbox{\tiny ww}}$).  
%=7.959\times 10^{-14}
In $U_{\mbox{\tiny ext}}^{\mbox{\tiny hb}}(x)$ (see eq.(2)), 
the $x$-dependence
of $\ep_s$ was approximated by a linear function increasing from its minimum value of 
$k^{\mbox{\mbox{\tiny }}}_{\mbox{\mbox{\tiny h}}}\ep_b$ 
at $x=\eta$ 
(with $k^{\mbox{\mbox{\tiny }}}_{\mbox{\mbox{\tiny h}}}=1.1$ corresponding to slightly 
enhanced hydrogen bonds$^{18,38}$ 
for molecules closest to the hydrophobic surface) to its maximum (bulk) value $\ep_b$ for 
$x\ge 2\eta$. This results in  $U_{\mbox{\tiny ext}}^{\mbox{\tiny hb}}(x)=
\ep_bn_b[(k^{\mbox{\mbox{\tiny }}}_{\mbox{\mbox{\tiny h}}} - 
(k^{\mbox{\mbox{\tiny }}}_{\mbox{\mbox{\tiny h}}} -1)(x-\eta)/\eta)n_s(x)/n_b - 1]$ for 
$\eta\le x <2\eta$ and $U_{\mbox{\tiny ext}}^{\mbox{\tiny hb}}(x)=0$ for $x\ge 2\eta$.

As clear from Figure 1, the hydrogen bonding contribution to the external potential has a
repulsive character unlike the conventional pairwise contribution that has an attractive
character (note also that the former dominates the latter in the most part of the
range $\eta< x<2\eta$). The repulsive character of $U_{\mbox{\tiny ext}}^{\mbox{\tiny hb}}$
arises because  that the total energy of hydrogen bonds per molecule near the hydrophobic
surface is smaller (in absolute value)  than in the bulk, which in turn is due to 
$n_s(x)\le n_b$ and $\ep_s(x)\approx\ep_b$ for any $x$.  

At a given temperature, the evaluation of the free energy of hydration is simpler in a grand canonical
esemble, i.e., by solving eq.(11), substituting the resulting equilibrium density profile in eq.(3), and
then calculating $\De \Omega_{\mbox{\tiny }}$ according to eq.(9). 
This procedure was applied to the hydration of 
an infinitely large flat hydrophobic surface in the model liduid water at temperature 
$T=293.15$ K and 
chemical potential $\mu=-11.5989$ $k_BT$ corresponding to the  
vapor-liquid equilibrium of the model fluid. 
The liquid state of the bulk water 
was ensured by imposing the appropriate boundary condition $\rho(x)\rightarrow
\rho_l$ as $x\rightarrow \infty$ onto eq.(11), 
with $\rho_l$ being the bulk liquid density. The densities
$\rho_v$ and $\rho_l$ of coexisting vapor and liquid, respectively, 
were found by solving a pair of 
equations expressing the conditions of phase equilibrium at a given $T$,
$$\left.\mu(\rho,T)\right|_{\rho=\rho_v}=\left.\mu(\rho,T)\right|_{\rho=\rho_l},\;\;\;\;\;\;\;\;
 \left.p(\rho,T)\right|_{\rho=\rho_v}=\left.p(\rho,T)\right|_{\rho=\rho_l},$$
%with the functions 
where$^{34}$ $\mu(\rho,T)=\mu_h(\rho,T)-\alpha\rho$ and 
$p(\rho,T)=p_h(\rho,T)-\frac1{2}\alpha\rho^2$ and 
positive constant $\alpha=-\int dr\,\phi_{\mbox{\tiny at}}(r_{})$. The pressure of a uniform
hard sphere fluid $p_h=p_h(\rho,T)$ is related to the corresponding chemical potential
$\mu_h$ via $\partial p_h/\partial \rho=\rho\partial \mu_h/\partial \rho$. In the
Carnahan-Starling approximation$^{34,35,41}$  
$$ p_h(\rho,T)=\rho k_BT \frac{1+\xi+\xi^2-\xi^3}{(1-\xi)^3}. $$

To clarify the effect of the two different contributions to $U_{\mbox{\tiny ext}}(x)$ on the water
density distribution near the hydrophobe, the density profiles were obtained$^{34}$ by solving  eq.(11)
with the  overall external potential a) including {\bf both} pairwise and hydrogen bond contributions
(i.e.,  $U_{\mbox{\tiny ext}}(x)=U_{\mbox{\tiny ext}}^{\mbox{\tiny pw}}(x)+ U_{\mbox{\tiny
ext}}^{\mbox{\tiny hb}}(x)$) and b) including {\bf only} the pairwise component (i.e., with 
$U_{\mbox{\tiny ext}}(x)=U_{\mbox{\tiny ext}}^{\mbox{\tiny pw}}(x)$). In both cases the ratio 
$k^{\mbox{\tiny }}_{\mbox{\tiny p}}$ was taken to be $2.1$.  It was clearly
demonstrated$^{36}$ that the hydrogen bond contribution to  the external potential plays a crucial role
in the formation of a thin, ``strong depletion" layer 
(of density much lower than liquid and 
of thickness of a molecular diameter, in agreement with previous suggestions$^{12,13}$) 
between liquid water and hydrophobic surface even for weakly
hydrophobic surfaces (with high $k^{\mbox{\tiny }}_{\mbox{\tiny p}}$). 
It was also shown that, as expected, even for a relatively strong hydrophobic surface (with
low $k^{\mbox{\tiny }}_{\mbox{\tiny p}}$)  the conventional contribution to the external
potential (due to pairwise interactions between a water molecule and those of the substrate)
cannot cause the formation of a vapor-like layer near the surface, although it does lead to
a weak decrease in the vicinal fluid denstity compared to the bulk one.  

Figure 2 presents the grand canonical free energy of hydrophobic  hydration 
$\De \Omega_{\mbox{\tiny }}$, expressed in units of $k_BT$ per $\eta^2$, as a function of the
energetic alteration ratio $k^{\mbox{\mbox{\tiny }}}_{\mbox{\mbox{\tiny h}}}$ of hydrogen
bonds (in the  SHL compared to the bulk),   at a constant ratio $k^{\mbox{\tiny
}}_{\mbox{\tiny p}}$  (characterizing the degree of the hydrophobicity of the surface).  Due
to the model character of the hydrophobic surface, we considered several values of 
$k^{\mbox{\tiny }}_{\mbox{\tiny p}}$ (each curve in Figure 2a corresponds to a constant 
$k^{\mbox{\tiny }}_{\mbox{\tiny p}}=2.0,2.1,2.2$ and $2.3$ from bottom to top). As clear, 
the hydrophobic hydration is hardly sensitive to the hydrogen bond energy alteration ratio, 
$k^{\mbox{\tiny }}_{\mbox{\tiny h}}$, but quite sensitive to the degree of hydrophobicity
of the surface,  $k^{\mbox{\tiny }}_{\mbox{\tiny p}}$. The latter effect is also
demonstrated in Figure 2b, where  the dependence of $\De \Omega_{\mbox{\tiny }}$ on
$k^{\mbox{\tiny }}_{\mbox{\tiny p}}$ is shown for the case where in eqs.(3) and (11) {\bf 
only} the pairwise component was included in  the external potential,  i.e., $U_{\mbox{\tiny 
ext}}(x)=U_{\mbox{\tiny ext}}^{\mbox{\tiny pw}}(x)$. 

The temperature effect on the strength of solvent-mediated part of hydrophobic attraction is 
demonstrated by Figure 3 that presents the Helmholtz free energy of hydrophobic hydration,  
$\De F_{\mbox{\tiny }}$,  and its energetic and entropic components,  $\Phi_{\mbox{\tiny E}}$
and $\Phi_{\mbox{\tiny S}}$,  as functions of $T$.  The solid curve represents $ \De
F_{\mbox{\tiny }}$ itself,  while the long-dashed and short-dashed curves are for 
$\Phi_{\mbox{\tiny E}}$ and $\Phi_{\mbox{\tiny S}}$, respectively. All the potentials are in
units of $k_BT$ per $\eta^2$.  The results in Figure 3a correspond to the 
overall external potential in eqs.(7) and (12) that includes  both the pairwise and hydrogen bond
contributions (i.e.,  $U_{\mbox{\tiny ext}}(x)=U_{\mbox{\tiny ext}}^{\mbox{\tiny pw}}(x)+
U_{\mbox{\tiny ext}}^{\mbox{\tiny hb}}(x)$),  whereas Figure 3b shows the results obtained by
solving  eq.(12) and (7) with {\bf only} the pairwise component included in the external
potential (i.e., with  $U_{\mbox{\tiny ext}}(x)=U_{\mbox{\tiny ext}}^{\mbox{\tiny pw}}(x)$). 
Both Figures 3a and 3b are for a hydrophobic surface with  $k^{\mbox{\tiny
}}_{\mbox{\tiny p}}=2.1$. As clear,  the water hydrogen bonding
plays an important role in the hydrophobic hydration by making the process
thermodynamically more unfavorable (i.e.,significantly  increasing the unfavorable free energy
of hydration). 

As shown in Figure 3, the combined PHB/DFT model predicts the free energy of hydration of  a
large  hydrophobic particle to decrease with increasing temperature and  suggests that the
hydration process is enthalpically unfavorable (i.e., the enthalpic contribution to the
hydration free energy is positive), but  entropically favorable (i.e., the entropic
contribution to the hydration free energy is negative), with the latter effect being dominant.
Currently, no experimental data on the thermodynamics of hydration of large hydrophobes are
available in literature.  However, the enthalpic impediment to the hydration of a large
hydrophobe is quite  expected$^{12,13,27-29}$ (being due to the breaking of vicinal hydrogen
bonds).  On the other hand, while the entropic enhancement of such hydration  seems somewhat
counter-intuitive, there are indirect experimental and simulational  indications of its
physical soundness. For example, ref.44 reported  experimental observation that
dissolving an argon molecule in hot liquid water leads to  an increase in entropy (i.e., the
hydration of an argon molecule is entropically favorable), although the transfer of the same
molecule into cold liquid water causes  a decrease in entropy (i.e., its hydration is
entropically unfavorable). These experiments are supported by a theoretical
model (the two-dimensional Mercedes-Benz model with one fitting parameter)$^{46}$  as well as
by the molecular dynamics simulations of SPC/E water model.$^{47}$ 

Furthermore, studying the lengthscale dependence of hydrophobic hydration (under various
thermodynamic conditions) by means of MD simulations of SPC/E water model, it was
demonstrated$^{23}$  that the hydration thermodynamics changes its character from ``entropy
dominated" to ``enthalpy dominated" near the crossover region as the length scale of a
hydrophobe increases. At $T=300$ K and pressure $-1000$ atm,  the crossover region was found
to be around $R=3$ nm ($R$ being the radius of a spherical  hydrophobe). As reported,$^{23}$ 
the hydration is  predominantly entropic ($\Phi_{\mbox{\tiny S}}>\Phi_{\mbox{\tiny E}}>0$), 
for  solutes of radii smaller than $3$ nm, whereas it is predominantly enthalpic
($\Phi_{\mbox{\tiny E}}>\Phi_{\mbox{\tiny S}}>0$) for  solutes of radii larger than $3$ nm.
As the radius of the solute increases, the enthalpic
contribution $\Phi_{\mbox{\tiny E}}$  increases whereas the entropic contribution $\Phi_{\mbox{\tiny
S}}$ decreases
and is expected to become negative ``for sufficiently large solutes" (see ref.23). 
Qualitatively, this can be interpreted as the result of breaking the tetrahedrally ordered
structure of  water-water hydrogen bond network by the foreign hydrophobic particle; breaking
the ordered structure of the hydrogen bond network is equivalent to increasing the disorder in
the system which leads to an increase in  its entropy. 

\section{Concluding remarks}

In order to clarify some aspects of the effect of water-water hydrogen bonding on the
thermodynamics of hydrophobic hydration, we have proposed a combination of our previously
developed probabilistic approach to water-water hydrogen bonding with the classical density
functional theory.   The latter allows one to accurately determine  the distribution of water
molecules in the vicinity of a hydrophobic particle and calculate the thermodynamic
quantities of hydrophobic hydration as well as their temperature dependence. The former
allows one to implement the effect of the hydrogen bonding ability of  water molecules on
their interaction with the hydrophobic surface into the DFT formalism. 

The hydrogen bond network of water molecules affects their interaction with the hydrophobic
surface  because the number and energy of hydrogen bonds that a water  molecule forms  in the
the surface differ from their bulk values.  Such an alteration gives rise to  a
short-range hydrogen bond contribution to the external potential field whereto a water
molecule is  subjected in that vicinity.  This contribution is a dominant component
of  the interactions of a water  molecule with the surface at distances between one and two 
hydrogen bond length.  As we previously showed,$^{34}$ it  plays a crucial role in the
formation of a thin depletion layer (of thickness of a molecular diameter and of very low
density, in agreement with previous suggestions$^{12,13}$) between liquid water and
hydrophobic surface.

The combined PHB/DFT approach to hydrophobic hydration predicts that the free energy of
hydration of a model hydrophobic surface in a model liquid water decreases with increasing
temperature in the range from 293 K to 333 K. It also corroborates the counter-intuitive
experimental, simulational, and theoretical observation  that under some thermodynamic 
conditions the hydrophobic hydration may be entropically favorable. As a possible explanation,
one can conjecture that the destruction of the  tetrahedrally ordered structure of 
water hydrogen bond network by a hydrophobic particle results in an
increased disorder in the system which leads to an increase in its entropy. 

{\em Acknowledgement} - 
{\small The author thanks Dr. E. Ruckenstein and Dr. G. Berim for many helpful discussions. 
%This work was supported by the National Science Foundation through  the
%grant CTS-0000548
}

\section*{References}
\begin{list}{}{\labelwidth 0cm \itemindent-\leftmargin} 
\item $(1)$ Pimental, C.G.; McCellan, A.L. {\it The Hydrogen Bond}; 
W.H.Freeman: San Francisco, 1960. 
%\item $(2)$  Huggins, 1971
\item $(2)$ P.Schuster, G.Zindel, and C. Sandorfy (eds), 
{\it The Hydrogen Bond: Recent Developments in
Theory and Experiments}, 3 vols (North Holland, Amsterdam, 1976).
\item $(3)$ a) Chaplin, M.F. In {\em Water of Life: The unique properties of H$_2$0}; 
Lynden-Bell, R.M.; Morris, S.C.; Barrow, J.D.; Finney, J.L.; Harper, C., Eds; CRC Press, Boca Raton,
2010; p.69;  
b) M.F.Chaplin, {\em Water Structure and Science}, 
(e-Book, http://www.lsbu.ac.uk/water/index.html).
\item $(4)$ Sharp, K.A. {\it Curr. Opin. Struct. Biol.} {\bf 1991}, {\it 1}, 171. 
\item $(5)$ Soda, K. {\it Adv. Biophys.} {\bf 1993}, {\it 29}, 1. 
\item $(6)$ Paulaitis, M.E.; Garde, S.; Ashbaugh, H.S.  
{\it Curr. Opin. Colloid Interface Sci.} {\bf 1996}, {\it 1}, 376.
\item $(7)$ Blokzijl, W.; Engberts, J.B.F.N. {\it Angew. Chem. Int. Ed. Engl.} {\bf 1993}, 
{\it 32}, 1545-1579.
\item $(8)$ Anfinsen, C.B. {\it Science} {\bf 1973}, {\it 181}, 223-230.
\item $(9)$ Ghelis, C.; Yan, J. {\it Protein Folding}; Academic Press: New York, 1982.
%\item $(6)$ T.E.Creighton,  (1984) in Proteins: Structure and Molecular Properties, W.H.Freeman, San 
%   	Francisco.
\item $(10)$ Kauzmann, W. {\it Adv.Prot.Chem.} {\bf 1959}, {\it 14}, 1-63.
\item $(11)$ Privalov, P.L. {\it Crit.Rev.Biochem.Mol.Biol.} 
{\bf 1990}, {\it 25}, 281-305.
\item $(12)$ Ball, P. {\it Chem.Rev.} {\bf 2008}, {\it 108}, 74-108.
\item $(13)$ Berne, B.J.; Weeks, J.D.; Zhou, R. {\it Annu.Rev.Phys.Chem.} {\bf 2009}, {\it 60}, 
85-103.
\item $(14)$ Frank, H.; Evans, M. {\it J.Chem.Phys.} {\bf 1945}, {\it 13}, 507-532.  
\item $(15)$ Stillinger, F.H. {\it J.Solut.Chem.} {\bf 1973}, {\it 2}, 141-158.
\item $(16)$ Pratt, L.R.; D.Chandler, D. {\it J.Chem.Phys.} {\bf 1977}, {\it 67}, 3683-3704.
%\item $(9)$ G.T.Barkema and B.Widom, J.Chem.Phys. {\bf 113}, 2349 (2000); 
\item $(17)$ M\"uller, {\it N. Accounts Chem.Res.} {\bf 1990}, {\it 23}, 23-28.
\item $(18)$ Lee, B.; Graziano, G. {\it J.Am.Chem.Soc.} {\bf 1996}, {\it 118}, 5163-5168.
\item $(19)$ Widom, B.; Bhimulaparam, P.; Koga, K. {\it Phys.Chem.Chem.Phys.} {\bf 2003}, {\it 5}, 
3085-3093.
\item $(20)$ Pratt, L.R. {\it Annu. Rev. Phys. Chem.} {\bf 2002}, {\it 53}, 409-436.
\item $(21)$ Lum, K.; Chandler, D.; Weeks, J.D. {\it J.Phys.Chem.B} {\bf 1999}, {\it 103}, 4570-4577.
\item $(22)$ Southall, N.T.; Dill, K.A. {\it J.Phys.Chem.B} {\bf 2000}, {\it 104}, 1326-1331.
\item $(23)$ Rajamani, S.; Truskett, T.M.; Garde, S. {\it Proc.Natl.Acad.Sci.USA} {\bf 2005}, 
{\it 102} 9475-9480.
\item $(24)$ Chandler, D. {\it Nature} {\bf 2005}, {\it 437}, 640-647.
\item $(25)$ Pangali, C.; Rao, M.; Berne, B.J. {\it J.Chem.Phys.} {\bf 1979}, {\it 71}, 2982-2990. 
\item $(26)$ Watanabe, K.; Andersen, H.C. {\it J.Phys.Chem.} {\bf 1986}, {\it 90}, 795-802.
\item $(27)$ Lee, C.Y.; McCammon, J.A.; Rossky, P.J. {\it J.Chem.Phys.} {\bf 1984}, {\it 80}, 4448-4455.
\item $(28)$ Choudhury, N.; Pettitt, B.M. {\it Mol.Simul.} {\bf 2005}, {\it 31}, 457-463.   
\item $(29)$ Choudhury, N.; Pettitt, B.M. {\it J.Am.Chem.Soc.} {\bf 2007}, {\it 129}, 4847-4852.
\item $(30)$ Evans, R. {\it Adv. Phys.} {\bf 1979}, {\it 28}, 143-200.
\item $(31)$ Evans, R. In {\it Fundamentals
of inhomogeneous fluids}; Henderson, D., Ed.; Marcel Dekker: New York, 1992. 
\item $(32)$ Sullivan, D.E. {\it Phys.Rev. B} {\bf 1979}, {\it 20}, 3991-4000. 
\item $(33)$ Tarazona, P.; Evans, R. {\it Mol.Phys.} {\bf 1983}, {\it 48}, 799-831.
\item $(34)$ Ruckenstein, E.; Djikaev, Y.S. {\it J. Phys. Chem. Lett.} {\bf 2011}, {\it 2},
1382-1386. 
\item $(35)$ Djikaev, Y.S.; Ruckenstein, E. {\it J.Chem.Phys.} {\bf 2010}, {\it 133}, 194105. 
\item $(36)$ Djikaev, Y.S.; E. Ruckenstein, Curr. Opin. Colloid Interface Sci. 
doi:10.1016/j.cocis.2010.10.002 (2010).
\item $(37)$ Meng, E.C.; Kollman, P.A. {\it J.Phys.Chem.} {\bf 1996}, {\it 110}, 11460-11470.
\item $(38)$ Silverstein, K.A.T.; Haymet, A.D.J.; Dill, K.A. {\it J.Chem.Phys.} {\bf 1999}, {\it 111}, 
8000-8009.
\item $(39)$ Djikaev, Y.S.; Ruckenstein, E. {\it J.Chem.Phys.} {\bf 2009}, {\it 130}, 124713. 
\item $(40)$ Tarazona, P. {\it Phys. Rev. A} {\bf 1985}, {\it 31}, 2672-2679.
\item $(41)$ Curtin, W.A.; Ashcroft, N.W. {\it Phys. Rev. A} {\bf 1985}, {\it 32}, 2909-2919.
\item $(42)$ Nakanishi, H.; Fisher, M.E. {\it Phys.Rev.Lett.} {\bf 1982}, {\it 49}, 1565-1568.
\item $(43)$ Lee, D.J.; Telo da Gama, M.M.; Gubbins, K.E. {\it J.Chem.Phys.} {\bf 1986}, {\it 85}, 
490-499.
\item $(44)$ Carnahan, N.F.; Starling, K.E. {\it J.Chem.Phys.} {\bf 1969}, {\it 51}, 635-636.
\item $(45)$ Weeks, J.D.; Chandler, D.; Anderson, H.C. {\it J.Chem.Phys.} {\bf 1971}, {\it 54}, 
5237-5247.
\item $(46)$ Xu, H.; Dill, K.A. {\it J.Chem.Phys.} {\bf 2005}, {\it 109}, 23611-23617.
\item $(47)$ Guillot, B.; Guissani, Y. {\it J.Chem.Phys.} {\bf 1993}, {\it 99}, 8075-8094.

\end{list}

\newpage 
\subsection*{Captions} 
to  Figures 1 to 3 of the manuscript {\sc
``The effect of water-water hydrogen bonding on hydrophobic hydration and its temperature 
dependence"
}  by {\bf Y. S. Djikaev} and  {\bf E. Ruckenstein}. 
\subsubsection*{}
\vspace{-0.8cm}   
Figure 1. The typical behaviour of the overall 
external potential $U_{\mbox{\tiny ext}}$ (exerted by a hydrophobic surface on water molecules in
its vicinity), shown as a thick dash-dotted curve, and its components, 
$U_{\mbox{\tiny ext}}^{\mbox{\tiny pw}}$ (lower thin solid curve) 
and $U_{\mbox{\tiny ext}}^{\mbox{\tiny hb}}$ (upper thin solid curve).
\vspace{0.3cm}\\ 
Figure 2. 
The grand canonical free energy of hydrophobic hydration 
$\De \Omega_{\mbox{\tiny }}$, 
expressed in units of $k_BT$ per $\eta^2$: a) as a function of 
the bond energy alteration ratio $k^{\mbox{\mbox{\tiny }}}_{\mbox{\mbox{\tiny h}}}$ 
at a constant $k^{\mbox{\tiny }}_{\mbox{\tiny p}}$. 
Different curves correspond to different degrees of hydrophobicity 
($k^{\mbox{\tiny }}_{\mbox{\tiny p}}=2.0,\,2.1,\,2.2$, and $2.3$ from bottom to top); 
b) as a function of 
$k^{\mbox{\tiny }}_{\mbox{\tiny p}}$ for the case where $U_{\mbox{\tiny
ext}}(x)=U_{\mbox{\tiny ext}}^{\mbox{\tiny pw}}(x)$.
\vspace{0.3cm}\\ 
Figure 3.
The Helmholtz free energy of hydrophobic hydration,  
$\De F_{\mbox{\tiny }}$,  and its energetic and entropic components, 
$\Phi_{\mbox{\tiny E}}$ and $\Phi_{\mbox{\tiny S}}$, 
as functions of $T$ for a hydrophobic surface with 
$k^{\mbox{\tiny }}_{\mbox{\tiny p}}=2.1$ and a) 
$U_{\mbox{\tiny ext}}(x)=U_{\mbox{\tiny ext}}^{\mbox{\tiny pw}}(x)$)  and  
b) $U_{\mbox{\tiny ext}}(x)=U_{\mbox{\tiny ext}}^{\mbox{\tiny pw}}(x)+
U_{\mbox{\tiny ext}}^{\mbox{\tiny hb}}(x)$. 
The solid curves represent $ \De F_{\mbox{\tiny }}$ itself, 
while the long-dashed and short-dashed curves are for 
$\Phi_{\mbox{\tiny E}}$ and $\Phi_{\mbox{\tiny S}}$, respectively.

\newpage
\begin{figure}[htp]
\begin{center}\vspace{1cm}
\includegraphics[width=8.3cm]{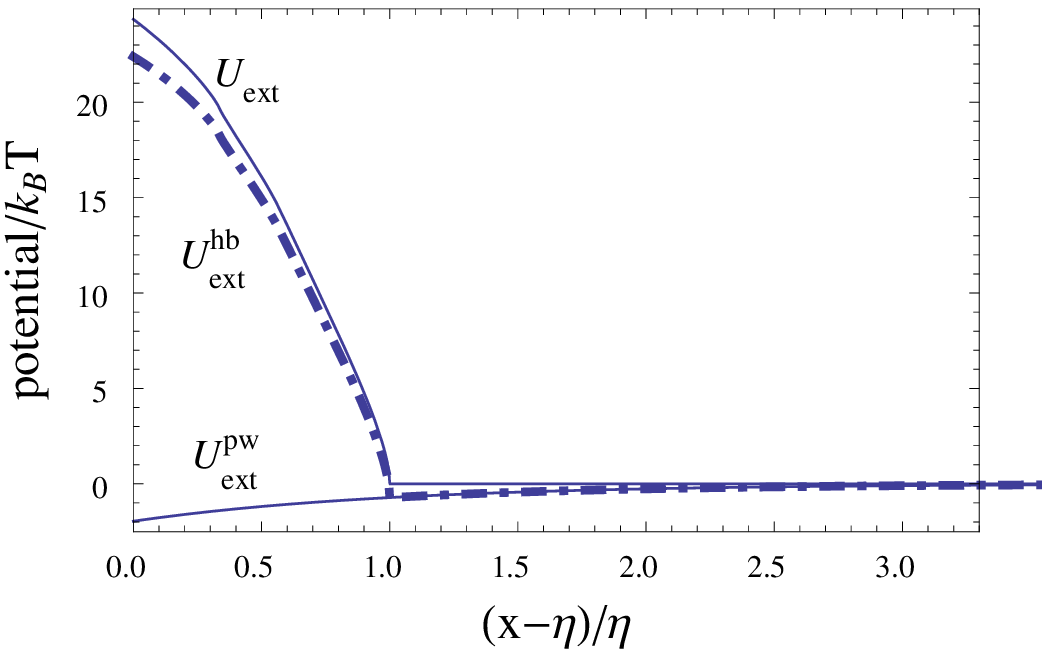}\\ [3.7cm]
\caption{\small }
\end{center}
\end{figure} 

\newpage
\begin{figure}[htp]\vspace{-1cm}
	      \begin{center}
$$
\begin{array}{c@{\hspace{0.3cm}}c} 
%	      \begin{flushright}
              \leavevmode
      	      \vspace{1.3cm}
	\leavevmode\hbox{a) \vspace{3cm}} &   
\includegraphics[width=8.3cm]{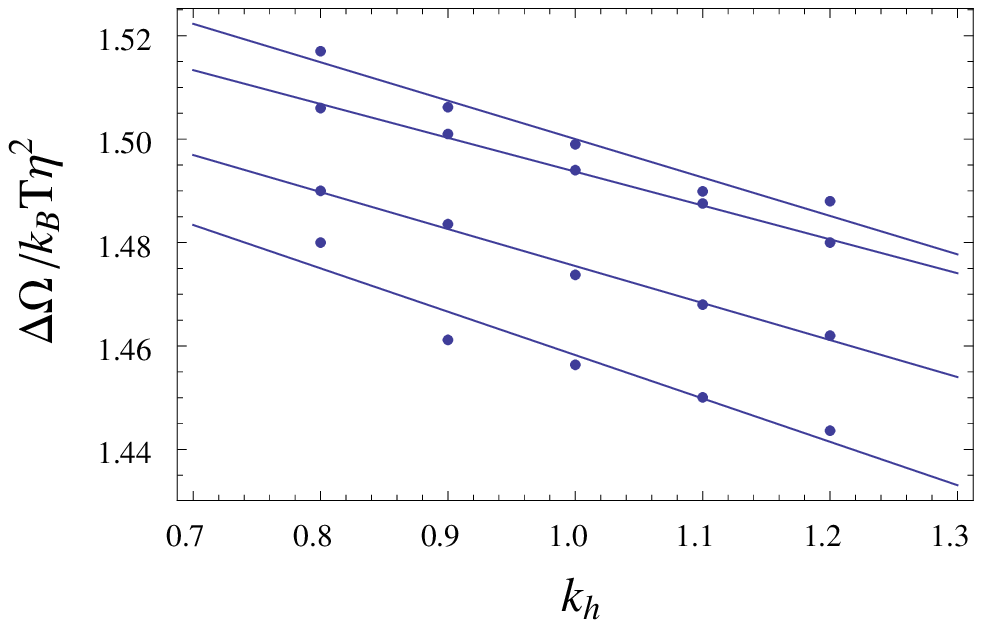}\\ [1.9cm] 
      	      \vspace{1.7cm}
	\leavevmode\hbox{b) \vspace{3cm}} &  
      	      \vspace{0.0cm}
\includegraphics[width=8.3cm]{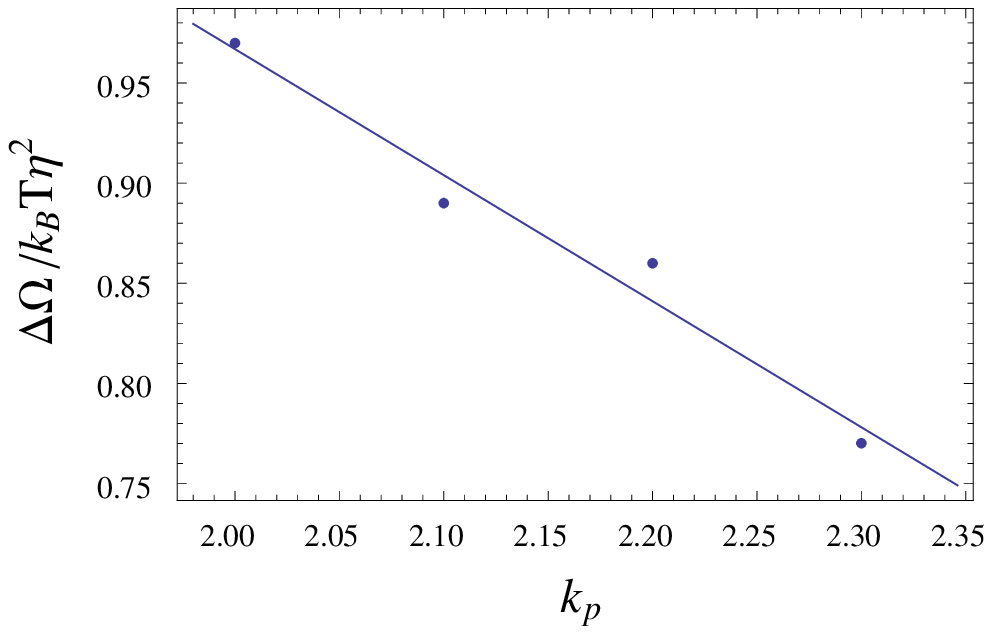}\\ [0.7cm] 
%              \end{flushright} 
%\mbox{\bf (aa)} & \mbox{\bf (bb)} 
\end{array}  
$$  

	      \end{center} 
            \caption{\small } 
\end{figure}

\newpage
\begin{figure}[htp]\vspace{-1cm}
	      \begin{center}
$$
\begin{array}{c@{\hspace{0.3cm}}c} 
%	      \begin{flushright}
              \leavevmode
      	      \vspace{1.3cm}
	\leavevmode\hbox{a) \vspace{3cm}} &   
\includegraphics[width=8.3cm]{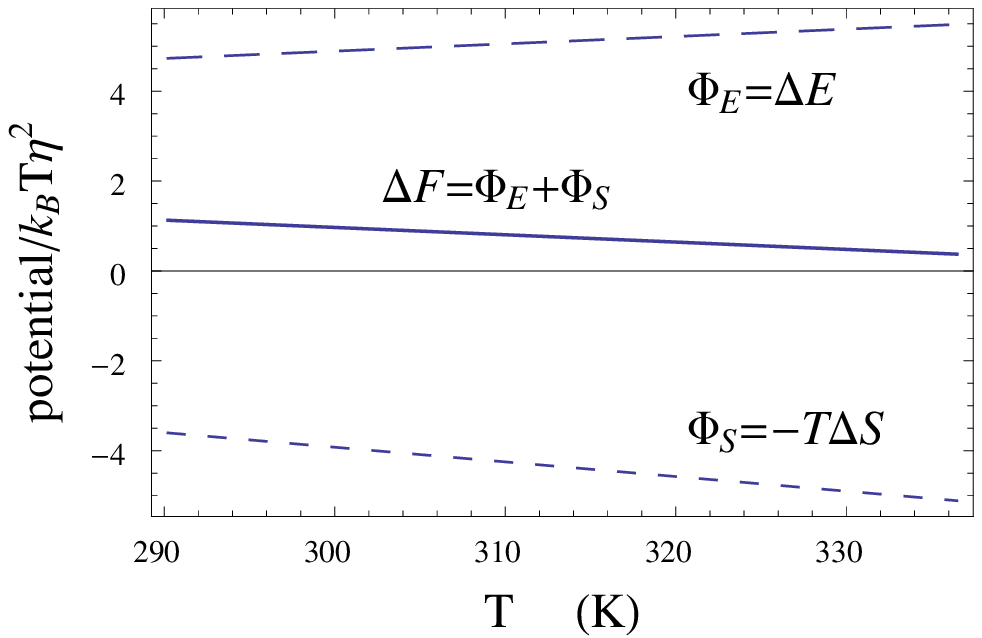}\\ [1.9cm] 
      	      \vspace{1.7cm}
	\leavevmode\hbox{b) \vspace{3cm}} &  
      	      \vspace{0.0cm}
\includegraphics[width=8.3cm]{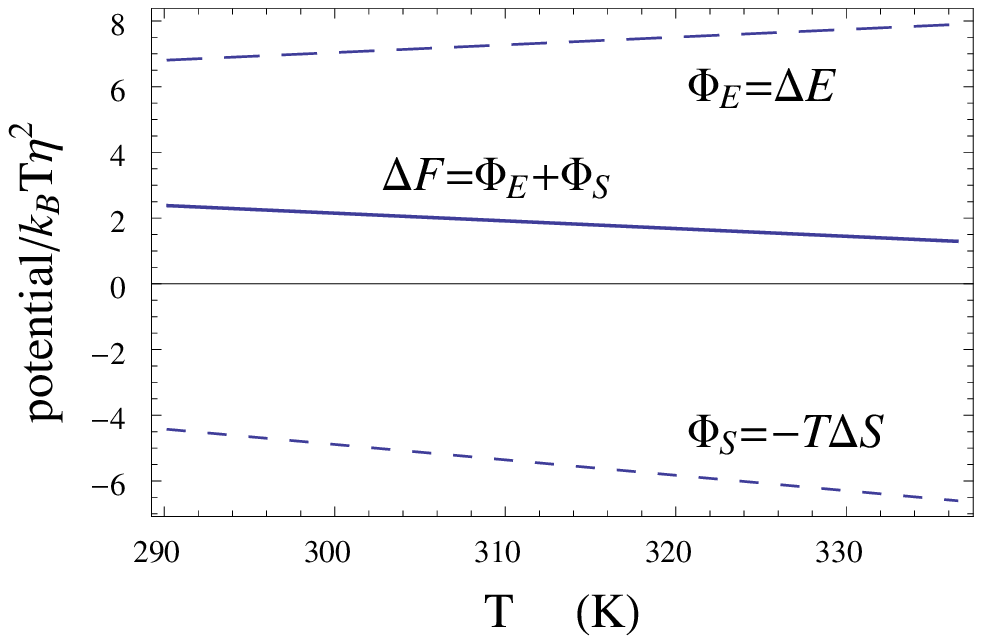}\\ [0.7cm] 
%              \end{flushright} 
%\mbox{\bf (aa)} & \mbox{\bf (bb)} 
\end{array}  
$$  

	      \end{center} 
            \caption{\small } 
\end{figure}

\end{document}